# Space-charge compensation experiments at IOTA ring

V. Shiltsev and M.Chung

*FNAL, PO Box 500, Batavia, IL 60510 USA*

**Abstract.** Space-charge effects belong to the category of the most long-standing issues in beam physics, and even today, after several decades of very active exploration and development of counter-measures, they still pose the most profound limitations on performance of high intensity proton accelerators. We briefly consider past experience in active compensation of these effects and present in detail the progress towards experimental studies of novel schemes of space-charge compensation at the Fermilab's IOTA ring.
.

## SPACE-CHARGE COMPENSATION : NEW PARADIGM FOR OLD PROBLEM

The effects of the space-charge forces belong to the category of the most long-standing issues in beam physics, and even today, after several decades of very active exploration and development of counter-measures, they set profound limitations on the performance of high intensity proton accelerators [1, 2]. They often manifest themselves in the form of beam loss, core emittance growth, and halo formation. In the past, several methods to compensate transverse SC effects were proposed and some of them attempted – see review of them in Ref. [3]. Below we outline only most recent proposals which are set for experimental tests at Fermilab

**Electron columns** [4] is an example of passive neutralization, when the SC force of a high-energy high-intensity proton beam is compensated by magnetized ionization electrons. The compensation condition reads $\eta \approx 1/\gamma^2_p$ . Optimum compensation requires that the transverse electron and beam distributions are matched. That could be achieved by confining the electrons transversely with strong solenoid fields to "columns" and using electrostatic electrodes to fine tune the charge density – see Fig.1. Strong magnetic field also stabilizes electron "column" motion and prevents coherent e-p instability. Simulations show significant reduction of SC induced emittance growth with only few "columns" occupying a small fraction $\eta$ of the ring circumference.

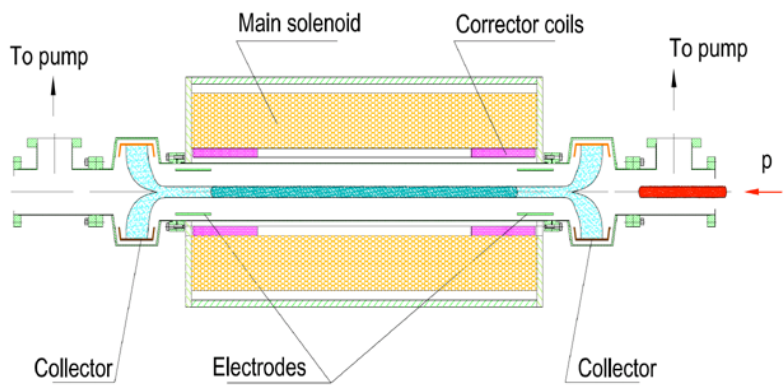

Figure 1: Schematic layout of an "electron column" for space-charge compensation [4].

**Electron lenses**, in which externally generated electron beam with matched transverse distribution collides with the proton beam inside a strong solenoid field, could also compensate the SC tune shift [5]. Assuming the total length of the lenses $L$, distributed around the ring, and an electron beams co-propagating with the proton beam, the electron current needed per lens is [6] $J_e = (B_f \kappa e c N_p/L)\, \beta_e/(\gamma^2_p (1 - \beta_p \beta_e))$, and for many accelerators of interest lays in the range of 1-10A for 10-40 keV electrons (here $\kappa$ denotes the degree of compensation, $B_f$ – proton bunching factor). These parameters are close to those of the operating Tevatron electron lenses. The SCC by lenses works better if the electron current is modulated to match longitudinal profile of proton bunches [7]. A practical method to achieve necessary time modulation of the electron focusing forces has been recently proposed in [8].

Recently proposed **fully nonlinear but integrable lattice** accelerators have promise to accommodate extraordinary large tune spreads in circulating beams without driving losses (resonance free optics) [9]. Comprehensive numerical studies of SC dynamics in the Integrable Optics rings have been started [10].

## SPACE-CHARGE COMPENSATION R&D AT IOTA RING

There are many challenges for the proposed space-charge compensation methods which call for experimental verification, including stability of the electron-proton system (transverse motion), (dynamic) matching of transverse proton-charge distribution, appropriate longitudinal compensation (for not-flat proton bunches), (dis)advantages of electron lenses vs electron columns, technology and practical implementation (in existing facilities), etc. A unique chance for carrying out the much needed dedicated studies is offered by new Fermilab's Accelerator R&D facility ASTA [11] – see Fig.2a.

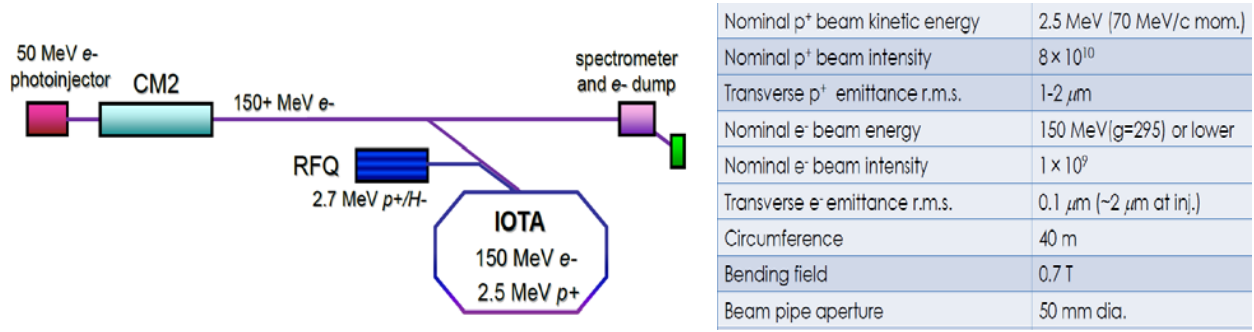

Figure 2: a) Schematic layout of ASTA R&D facility and b) IOTA ring parameters.

The Advanced Superconducting Test Accelerator (ASTA) at Fermilab incorporates a superconducting radiofrequency (SRF) linac coupled to a photoinjector and small-circumference storage ring capable of storing electrons or protons. ASTA will establish a unique resource for R&D towards Energy Frontier facilities and a test-bed for SRF accelerators and high-brightness beam applications. The unique features of ASTA include: (1) a high repetition-rate, (2) one of the highest peak and average brightness within the U.S., (3) a GeV-scale beam energy, (4) an extremely stable beam, (5) the availability of SRF and high-quality beams together, and (6) a storage ring called IOTA (Integrable Optics Test Accelerator – see parameters in Fig.2b) capable of supporting a broad range of ring-based advanced beam dynamics experiments [12].

The experiments planned to be carried out at the IOTA ring include the initial set which requires well-qualified narrow 150 MeV $e$- beam: Integrable Optics test with non-linear magnets, Integrable Optics test with e-lens(es), optical stochastic cooling Test, electron quantum wavefunction size, etc; which will be followed by the space-charge effects studies and compensation experiments with 2.5 MeV protons and/or $H$-: i) SC modes and dynamics in the ring with and without integrable optics, ii) SC compensation with e-columns; iii) SC compensation with e-lenses, iv) proton and H- halo and stripping, beam and halo diagnostics, etc. Below we present numerical simulations for one of the IOTA SC-compensation experiments – namely, the accumulation of charges in electron columns.

# SIMULATIONS OF THE IOTA ELECTRON COLUMN

Simulation of the plasma formation and trapping in the *e*-column have been carried out with use of WARP 3D code [13]. High intensity high brightness proton beam (2.5MeV kinetic energy, 8 mA of average beam current, RMS beam size of 5.5 mm, beam distribution is uniform in longitudinal direction and Gaussian in transverse direction, zero thermal spread) is injected into the IOTA ring equipped with a 1 m long electron column. Protons ionize the residual $H_2$ gas (pressure of 10-3 Torr for initial test, neutralization time of about 0.9 ms) and dynamics of ionization electrons and ions in *E* and *B* fields simulated while primary proton beam is considered stable. One can change solenoidal magnetic field *B* (typically, 0 to 1 T), gas pressure, voltages on the electrodes (typically, 0 to -4 kV), and geometry of the e-column such as length, number and diameters of the trapping electrodes(typically, 10 cm). The time step in simulations was 15 ps that is less then cyclotron period, each macroparticle represented some 150,000 real particles. The main reactions of interest were $p+H_2 \rightarrow p+H^+_2+e$ and $e+H_2 \rightarrow H^+_2+2e$ [14]. Simulations were carried out at the NERSC facility and took few weeks.

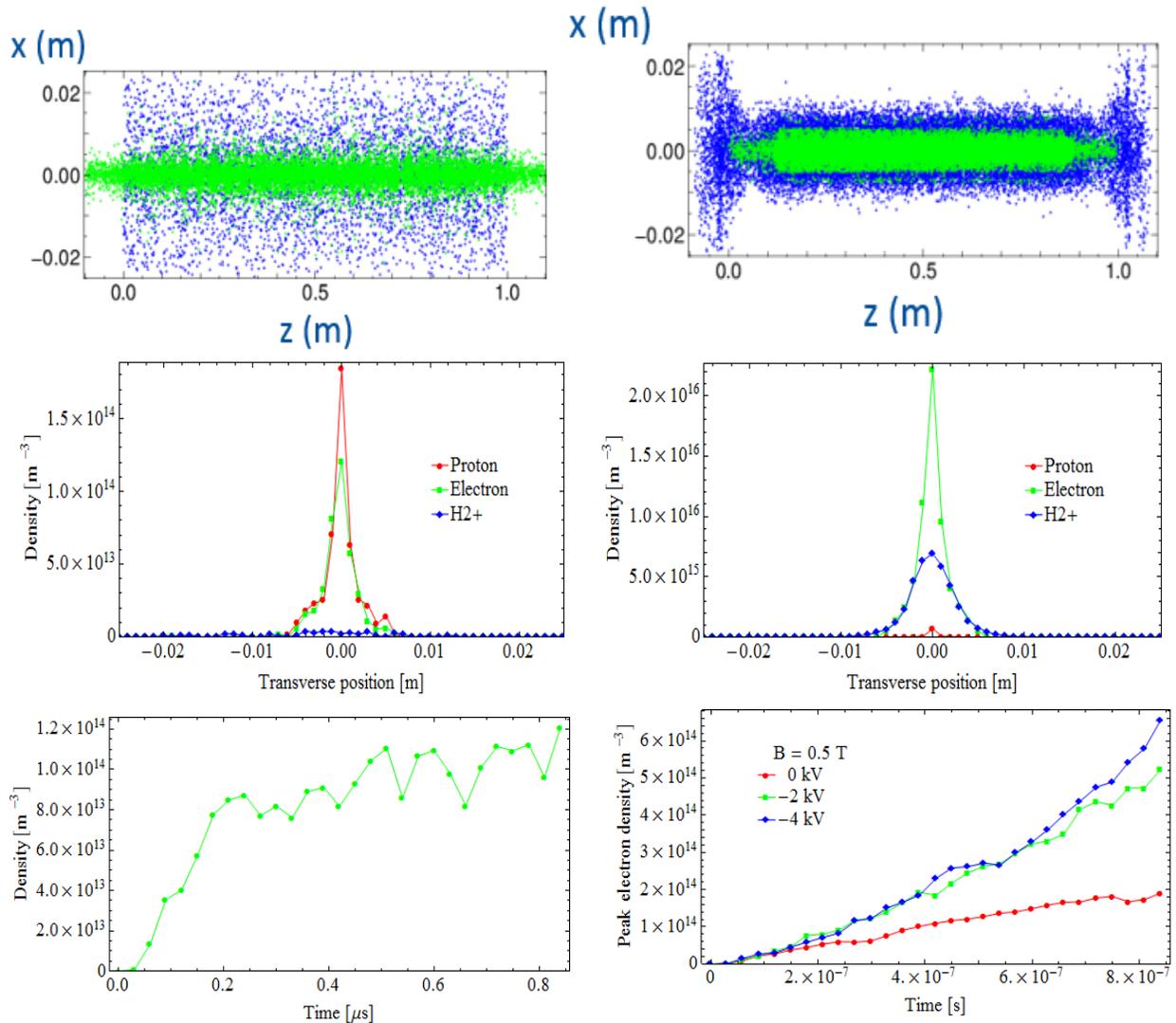

Figure 3: WARP 3D simulation of the plasma formation and trapping in the e-column: on the left - without magnetic field and no trapping voltage; on the right – with 0.5 T magnetic field and upto -4kV trapping voltage. Top to bottom: spatial distribution of electrons (green dots) and $H^+_2$ ions; transverse distribution of the proton, ion and electron charges, accumulation of electron charge density in the center of the column.

Main results of the studies can be summarized as follows : i) without longitudinal magnetic field $B=0$ and no trapping voltage $U=0$ kV ionization electrons accumulate in the proton beam, concentration is low (~1/2 of $p+$), the size of electron cloud is about that of $p+$; the $H_2+$ ions form a broad cloud that has only few % of the total charge; ii) without longitudinal magnetic field $B=0$ but with trapping voltage of $U=-2, -4kV$, the ionization electrons accumulate in between the electrodes and $H_2+$ ions accumulate near the electrodes, but the electron density is still low (~1/2 of $p+$), while the size about that of p+; iii) longitudinal magnetic field of $B=0.5$T or 1T drastically changes the situation: (with trapping voltage of $U=-2, -4kV$) the transverse diffusion of ionization electrons and their escape rate become very low, they effectively accumulate in between the electrodes and significantly overcompensate the primary $p+$ space-charge by a factor of 2-3, $H_2+$ ions are still present at the level of few %. Accumulation of the electrons in the presence of magnetic field continued beyond the neutralization time and the system exhibits a surprising degree of overcompensation - the phenomenon that needs further studies.

## DISCUSSION AND CONCLUSIONS

Progress of the Intensity Frontier accelerator based HEP is hindered by fundamental beam physics phenomena such as space-charge effects, beam halo formation, particle losses, transverse and longitudinal instabilities, beam loading, inefficiencies of beam injection and extraction, etc. The IOTA/ASTA facility at Fermilab is being built as a unique test-bed for transformational R&D towards the next generation high-intensity proton facilities. The experimental accelerator R&D at the IOTA ring with protons and electrons, augmented with corresponding modeling and design efforts will lay foundation for novel design concepts allowing substantial increase of the proton flux available for HEP research with Fermilab accelerators to multi-MW beam power levels at very low cost. The facility will also become the focal point of a collaboration of universities, National and international partners. The main goals of the IOTA/ASTA teams are: 1) construct and commission the IOTA storage ring and its proton and electron injectors, and establish reliable and time-effective operation of the facility for accelerator research program; 3) carry out transformative beam dynamics experiments such as *integrable optics* with non-linear magnets and with electron lenses, and *space-charge compensation* with electron lenses and electron columns.

There are many challenges for the proposed space-charge compensation methods which call for experimental verification, including stability of the electron-proton system (transverse motion), (dynamic) matching of transverse p-charge distribution, appropriate longitudinal compensation (for not-flat proton bunches), (dis)advantages of electron lenses vs electron columns, technology and practical implementation (in existing facilities), etc. The IOTA ring offers a unique opportunity for the much needed dedicated experimental studies.

We have started the program of modeling the beam dynamics effects associated with accumulations of neutralizing electrons in the electron column device for space-charge compensation. WARP-3D simulations of processes in the electron column show that longitudinal magnetic field of about of less than 1T leads to accumulation of significant number of electrons (overcompensation) with transverse distribution close to that of the primary proton beam. Further modeling should include realistic circulation and dynamics of the primary beam of protons (external focusing and multiple passages thru the electron column)

## ACKNOWLEDGMENTS

The authors wishes to thank E.Prebys and A.Valishev for helpful discussions on the practical aspects of the space-charge compensation experiments at the IOTA ring. Fermilab is operated by Fermi Research Alliance, LLC under Contract No. DE-AC02-07CH11359 with the United States Department of Energy.